# Effect of electron and hole doping on the superconducting and normal state properties of $MgB_2$

S.Jemima Balaselvi, A.Bharathi, V.Sankara Sastry, G.L.N.Reddy and Y.Hariharan

Materials Science Division, Indira Gandhi Center for Atomic Research, *Kalpakkam – 603 102*.

*Abstract*

*The variation of $T_C$ in $MgB_2$ has been systematically investigated for monovalent cation substitution in $Mg_{1-x}Li_xB_2$ and in $Mg_{1-x}Cu_xB_2$ and simultaneous cation and Carbon substitution in $Mg_{0.80}Li_{0.20}B_{2-x}C_x$ and $Mg_{0.95}Cu_{0.05}B_{2-x}C_x$ using dc resistivity and ac susceptibility techniques. $T_C$ seems to be uniquely determined by the electron count in the sample, being constant at the $MgB_2$ value for electron counts lower than $MgB_2$ but rapidly decreasing for larger electron counts. The $\theta_D$ extracted from fitting normal state resistivity to the Bloch-Gruneisen formula shows no systematics with $T_C$ while the residual resistivity versus $T_C$ indicates that interband/intraband scattering is affected by the substitutions.*

## INTRODUCTION

The origin of the unusually large $T_C$ in $MgB_2$[1], is thought to arise from double gap superconductivity, due to coupling of phonons with electrons in the $\sigma$ and $\pi$ bands at the Fermi surface[2]. Chemical substitutions in $MgB_2$[3,4] have resulted in a decrease in $T_C$. There have been theoretical predictions of an increase in $T_C$ by introducing a B-C network and hole doping by substitution of Mg with Cu[5] and Li[6]. In an attempt to verify these predictions, a series of samples with Li and Cu substitution at Mg site and C substitution at B site along with 20%Li and 5% Cu substitution at Mg site were synthesized and the measured parameters viz., $T_C$, residual resistivity, the residual resistivity ratio, $\rho(300K)/\rho(40K)$ and the Debye temperature, obtained from the normal state resistivity are compared with that in $MgB_{2-x}C_x$[4].

## EXPERIMENTAL

Samples of composition $Mg_{1-x}Li_xB_2$ $Mg_{1-x}Cu_xB_2$ $Mg_{0.80}Li_{0.20}B_{2-x}C_x$ and $Mg_{0.95}Cu_{0.05}B_{2-x}C_x$ for various x were prepared by the standard solid-vapour technique[4]. All samples were characterized by x-ray diffraction for phase purity and lattice parameter variations. Measurement of $\chi(T)$ in 300K-4.2K range were done using an ac mutual inductance technique and $\rho(T)$ in the standard four probe geometry.

## RESULTS AND DISCUSSION

From the XRD data it is clear that the Li solubility in $MgB_2$ is 30% (x=0.3), while that of Cu is only 5% (x=0.05). Carbon substitution upto a fraction of x=0.30 in the cation substituted samples are all found to be phase pure. The lattice parameters remain unchanged with cationic substitutions, resulting in no change in volume, whereas with carbon substitutions the lattice parameter along 'a' decreases monotonically with increase in C fraction with a corresponding decrease in the lattice volume.

$T_C$ variation upon substitutions as identified by 90% diamagnetic signal and zero resistivity is shown in Fig 1(a) and Fig 1(b), respectively. The $T_C$ decrease obtained from both techniques are seen to be very similar in all the series, studied.

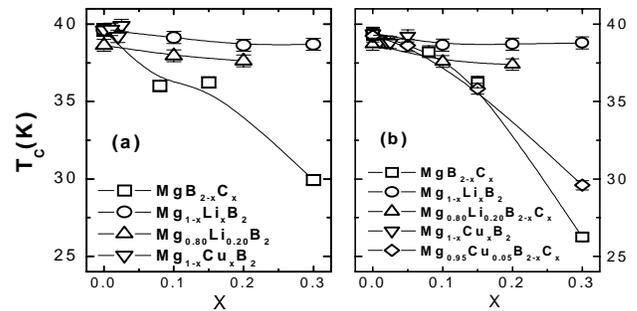

**Fig 1**: $T_C$ variation with substitution (a) from diamagnetic susceptibility and (b) zero resistivity

With cation substitution viz., $Mg_{1-x}Cu_xB_2$ and $Mg_{1-x}Li_xB_2$ a constancy in $T_C$ is observed while a decrease in $T_C$ is observed for the various C substitution. The extent of decrease in $T_C$ is seen to depend on the concentration of substituting cation, with a decrease of 14K, 10K and 1K being observed in $MgB_{2-x}C_x$ $Mg_{0.95}Cu_{0.05}B_{2-x}C_x$ and $Mg_{0.80}Li_{0.20}B_{2-x}C_x$ respectively. We define $N_{excess}=x-y$, where x denotes the carbon fraction and y denotes the cation fraction for the compound with general formula $Mg_{1-y}M_yB_{2-x}C_x$ (M=Li,Cu). Therefore, $N_{excess}$ is zero for $MgB_2$, positive for electron doped $MgB_2$ and negative for hole doped $MgB_2$. A plot of $N_{excess}$ against $T_C$ is shown in Fig 2 from which it is apparent that the $T_C$ remains almost constant with increase in the hole concentration while it decreases with increase in electron concentration. It is interesting to point out that Fig 2 holds a striking resemblance to the hole DOS versus energy curve obtained from band structure calculations[7], in which it was surmised that $T_C$ will not increase with increase in hole concentration whereas it would decrease with electron doping, being large for electron additions of ~0.2/formula unit

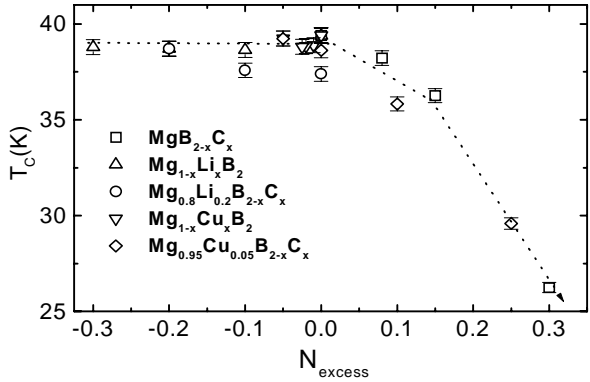

**Fig 2**: Plot of $N_{excess}$ versus $T_C$

Fitting the normal state resistivity to the Bloch-Gruneisen form

$$\rho(T) = \rho_0 + \rho_1 T^2 + K\,(m-1)\theta_D(T/\theta_D)^m * \int_0^{\theta_D/T} dz \frac{z^m}{(1-e^{-z})(e^z-1)},$$

with m=5, results in the extraction of $\rho_0$, $\rho_1$ and $\theta_D$. The value of $\rho_1$ was ~$10^{-10}$ indicating negligible electron scattering contribution. From $\theta_D$ and $T_C$, $\lambda_{MCM}$ is estimated using the McMillan equation with $\mu^*$=0.15. The variation of $\theta_D$ with $T_C$ for the different series is shown in Fig.3a. With Li and Cu substitution and with carbon substitution in $Mg_{0.95}Cu_{0.05}B_{2-x}C_x$, $\theta_D$ remains almost constant. In $MgB_{2-x}C_x$ and $Mg_{0.80}Li_{0.20}B_{2-x}C_x$ a decrease in $\theta_D$ larger than that expected from mass increase due to carbon substitution is observed. Fig. 3a also indicates the absence of any correlation between $\theta_D$ and $T_C$.

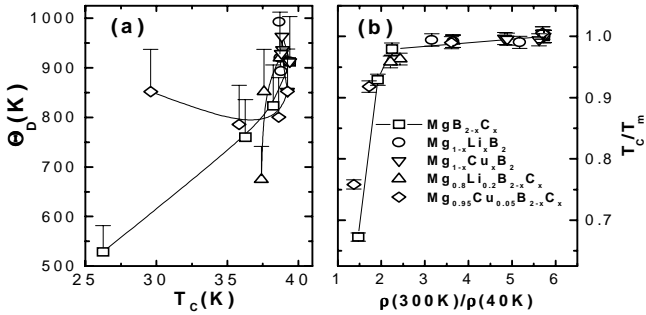

**Fig 3**: (a) Plot of $T_C$ versus $\Theta_D$ (b) Testardi correlation

The calculated $\lambda_{MCM}$ values for all samples fall in the range of 0.7 and 1.0 in agreement with reported values, in $MgB_2$. In Fig.3b is shown the Testardi plot of $\rho(300K)/\rho(40K)$ versus $T_C/T_m$, ($T_m=T_C$ in $MgB_2$) for all samples, suggestive of phonon mediated superconductivity in $MgB_2$ and related systems.

In Fig.4 the plot of $\rho(40K)$ versus $T_C$ is shown. It can be seen from the figure that $T_C$ versus $\rho(40K)$ is flat for Cu and Li substitutions (cf. inset Fig.4), whereas it shows a strong variation with carbon substitution. The steepest variation is seen for carbon substitution in which Cu has been substituted at the Mg site. A comparison of this data with recent theoretical calculations[8], that opines on the robustness of $T_C$ in $MgB_2$ despite large variations in $\rho_0$, suggests that intraband scattering in the $\sigma$ and $\pi$ bands is dominant in the Cu and Li substituted samples, whereas substitution of C results in an increase in the interband scattering, which is most pronounced in C doping of the Cu cation substituted samples. This can be understood as follows: C substitution could increase the $\pi$ character of the B-B bonds due to the extra $p_z$ electron and may thus contribute to the enhanced $\sigma$-$\pi$ hybridization. This hybridisation is possibly re-inforced due to 3d electrons in the Cu substituted samples. A quantitative understanding of the data shown in Fig.4 would be possible with band structure calculations similar to that in Ref.[8].

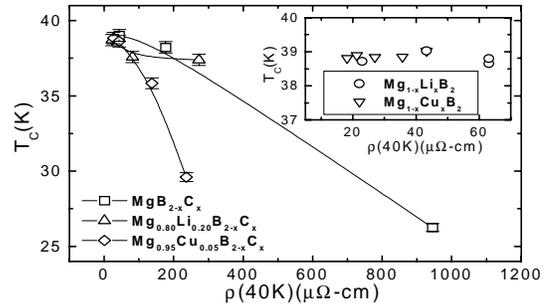

**Fig 4**: Plot of $\rho(40K)$ versus $T_C$

## SUMMARY AND CONCLUSION

$T_C$ shows no change with hole doping, while it decreases with electron doping. Testardi correlation and the fit of normal state resistivity to the Bloch-Gruneisen equation suggests phonon mediated transport and superconductivity in the system. Variation of $\rho(40K)$ with $T_C$ throws light on the relative magnitudes of intraband/interband scattering which seems to be altered by chemical substitutions.